\documentstyle[12pt]{article}
\def\beq{\begin{equation}}
\def\eeq{\end{equation}}
\def\bea{\begin{eqnarray}}
\def\eea{\end{eqnarray}}
\begin{document}
%\preprint{}

\begin{center}
{\Large  Expectation value of the axial-vector current in the
external electromagnetic field.}

{\bf Ara N. Ioannisian}

{\it   Yerevan Physics Institute, Alikhanian Br.  2, Yerevan-36,
Armenia}
\end{center}

\begin{abstract}
We are calculated the expectation value of the axial-vector current induced by the vacuum polarization effect of the Dirac field in constant external electromagnetic field.
In calculations we use Schwinger's proper time method. 
The effective Lagrangian has very simple Lorenz
invariant form. Along with the anomaly term, it also contains two  Lorenz
invariant terms. 
The result is compared with our previous
calculation of the photon - Z boson mixing in the magnetic field.
\end{abstract}

In order to get the expectation value of the axial-vector current in  constant electromagnetic
field we use Schwinger's  proper time method
\cite{Schwinger:1951nm}. In our calculations we closely follow and use results from \cite{Schwinger:1951nm} and \cite{Adler:1971wn}.

Green function of the Dirac field (with mass $m$ ) in the external electromagnetic
field $A_\mu$ and in the axial vector field $A_\mu^5$ (with couplings $e$
 and $g_a$ respectively) is

\bea
G &=& \frac{1}{\gamma(-i\partial-e A -g_a \gamma_5 A^5)+m}
\\
&=&\frac{1}{\gamma P+m}
\\
&=&i(-\gamma P+m) \int_0^{\infty}ds e^{-i(m^2-(\gamma P)^2)s} \ ,
\\
%A^5_\mu&=&\frac{G_F}{\sqrt{2}}\bar{\nu} \gamma_{\mu}(1-\gamma_5)\nu
%\\
g_{\mu\nu} &=& diag(-1,1,1,1) \ .
 \eea

The variation of the action integral is
 \bea
\label{1}
&&\delta S = \int dx
\left(ie\delta A_{\mu} Tr \gamma_{\mu}
(x{^\prime} \mid G  \mid x^{\prime\prime} )
+ig_a\delta A_{\mu}^5 Tr\gamma_{\mu} \gamma_5
(x{^\prime} \mid G \mid x^{\prime \prime}) \right)_
{x^\prime,x^{\prime \prime} \to x}
\\
&&=\int dx \delta {\cal L}  .
 \eea
In pure electrodynamics (without axial-vector field) we have only the first term under the integral in eq. (\ref{1}). Our task is to calculate the second term under the integral in eq. (\ref{1}) (which is just $\delta A_\mu^5 <J_\mu^5> $ where $ <J_\mu^5> $  is expectation value of the axial-vector field).

${\cal L}$ is the
Effective Lagrangian and we write
 \beq
{\cal L} = \frac{i}{2}\int_0^{\infty}\frac{ds}{s}e^{-i m^2 s}
\lim_{x^\prime,x^{\prime \prime} \to x} Tr(x{^\prime} | U(s) |
x^{\prime \prime} ) \ .
 \eeq
Here we use following notations 
 \bea
&&U(s)=e^{-i{\cal H} s} \ ,
\\
&&{\cal H} = -(\gamma P)^2
\\
&&{\cal H} = {\cal H}_0 +{\cal H}_I \ ,
\\
&&{\cal H}_0 = -(\gamma(-i\partial-e A))^2 = -(\gamma \Pi)^2 =
\Pi^2-\frac{1}{2}e \sigma_{\mu \nu} F_{\mu \nu} \ ,
\\
&&\Pi_\mu = -i\partial_\mu -eA_\mu \ ,
\\
&&F_{\mu \nu}=\partial_\mu A_\nu -\partial_\nu A_\mu = {1 \over
ie} [\Pi_\mu, \Pi_\nu] \ ,
\\
&&{\cal H}_I = g_a \gamma_5 [A_{\mu}^5,\Pi_{\mu}] + ig_a
\gamma_5\sigma_{\mu \nu} \{A_{\mu}^5,\Pi_{\nu}\} - g_a^2 {A^5}^2 \
 .
 \eea

$U(s)$ is the operator which develops the system with the
'Hamiltonian' ${\cal H}$ from the 'time' s=0 to the 'time' s.
${\cal H}_0$ is the 'Hamiltonian' of the pure electromagnetic
field and ${\cal H}_I$ is the 'Hamiltonian' of the rest which
contains axial vector field.

Applying to the identity
    \beq e^{A+B} = e^{A} T e^{ \int_0^1 dt
        e^{-A t}B e^{A t}},
    \eeq
where $T$ means the chronological order of the product, we can
write
    \bea
    U(s)&=& U_0(s)U_I(s)  \ , \\
    U_I(s)&=&T e^
    {-ig_a \gamma_5\int_0^s du \left( [A_{\mu}^5(u),\Pi_{\mu}(u)]+
    i\tilde\sigma_{\mu \nu} \{A_{\mu}^5(u),\Pi_{\nu}(u)\}
    -g_a A_{\mu}^5(u)^2\right)}  \ .
    \eea
Here
 \beq
\tilde\sigma_{\mu \nu} = e^{-\frac{i}{2}\sigma F eu}\sigma_{\mu
\nu} e^{\frac{i}{2}\sigma F eu} \ .
 \eeq

Keeping only linear term by axial vector field we get \beq
\label{qmm1} U(s)=U_0(s)\left(-ig_a \gamma_5\int_0^s du \left(
[A_{\mu}^5(u),\Pi_{\mu}(u)]+ i\tilde\sigma_{\mu \nu}
\{A_{\mu}^5(u),\Pi_{\nu}(u)\} \right) \right) \ .
 \eeq

The quantum mechanical problem can be formulated as

\beq
\label{qm}
(x^{\prime} | U(s) | x^{\prime \prime} ) = (x^{\prime}(s) | U_I(s) |
x^{\prime\prime} (0)) \ ,
 \eeq
where
 \bea
&&(x{^\prime} | U_0(s) \equiv (x^{\prime}(s) |  \ ,
\\
&& | x^{\prime \prime}) \equiv  x^{\prime\prime}(0) ) \ ,
\\
&&\Pi(0) \equiv \Pi \ ,
\\
&&\Pi(s) \equiv  U^{-1}(s)\Pi(0)U(s) \ ,
\\
&&x^{\prime \prime}(0) \equiv x^{\prime \prime} \ ,
\\
&&x^\prime(s) \equiv U^{-1}(s) x^\prime(0)U(s) \ .
 \eea

As Schwinger has shown
 \bea
 &&(x^{\prime}(s) |x^{\prime\prime}(0))
= \frac{-i}{s^2(4\pi)^2} \Phi(x^{\prime},x^{\prime \prime})
e^{(x^{\prime}-x^{\prime \prime}) E(s)(x^{\prime}-x^{\prime
\prime})} e^{-l(s)}e^{\frac{i}{2}e\sigma F s} \ ,
\\
&&\Phi(x^{\prime},x^{\prime \prime}) =e^{i e \int_{x^{\prime
\prime}}^{x^\prime} dx A(x)} \ ,
\\
&&E(s)=\frac{i}{4}eF\coth (eFs) \ ,
\\
&&l(s)=\frac{1}{2}Tr \ln {[}(eFs)^{-1}\sinh (eFs){]} \ .
 \eea

To evaluate the matrix element [\ref{qm}] we use the Schwinger's
results
 \bea
&&\Pi(u) = R(u) I^{-1}(s){[}x^{\prime}(s)-x^{\prime \prime}(0){]}
\ ,
\\
&&R(u) = e^{2eFu} \ ,
\\
&&I(s) = (eF)^{-1}(e^{2 e F s}-1) \ ,
\\
\label{com}
&&{[}x(s),x(0){]} = -i I(s) \ ,
\\
&& {[}x_{\mu},\Pi_\nu{]} = i g_{\mu \nu} \ ,
\\
&&\sigma_{\mu \nu} = \frac{i}{2} {[}\gamma_\mu,\gamma_\nu{]} \ .
 \eea

At low energy or in slowly varying axial vector field
approximation we can write
    \beq
    A^5_\mu (x(u))  \simeq A^5_\mu +\frac{\partial
    A^5_\mu}{\partial x_\nu} x_\nu(u)= A^5_\mu  +ik_\nu A^5_\mu
    x_\nu(u) \ ,
    \eeq
    where
    \beq
    x(u)=[1-I(u)I^{-1}(s)]x(0)+I(u)I^{-1}(s)x(s) \ .
    \eeq

In Eq. [\ref{qmm1}] we bring all factors $x^\prime(s)$ to the left
and all factors $x^{\prime \prime}(0)$ to the right  ( using
commutative relation [\ref{com}] ), where they act on the left-
and right-hand states to give c-numbers
 \beq
 x^{\prime \prime}(0)| x^{\prime \prime}(0)) = x^{\prime \prime} | x^{\prime
\prime}(0)) \ , \ \ \ \ \ \  \ \ \
 (x^{\prime} (s)| x^{\prime}(s)
= (x^{\prime} (s)| x^{\prime}  \ .
 \eeq

This procedure gives a c-number expression multiplied by the
function $(x^{\prime}(s) |x^{\prime\prime} (0))$.
\bea
\nonumber
&&(x^{\prime}(s) | U_I(s) | x^{\prime\prime} (0))=
(x^{\prime}(s) | x^{\prime\prime}(0)) \cdot
\\
&& \hspace{5 mm} \cdot (-ig_a\gamma_5 A_\mu^5)\int_0^s du \left[
\left(-N_{\mu\nu} - i\tilde\sigma_{\mu \lambda}D_{\lambda \nu}
\right)k_\nu + i\tilde\sigma_{\mu \lambda}O_{\lambda \nu}
(x^{\prime} - x^{\prime\prime})_\nu\right] \ ,
 \eea
where
 \bea
 &&N_{\mu \nu}= g_{\mu \nu} \ ,
\\
&&D_{\lambda \nu} = e^{2eFu}-\coth eFs ( e^{2eFu}-1) \ ,
\\
&&{\cal D}_{\lambda \nu} \equiv \int_0^s du D_{\nu \lambda} =
{[}-\frac{1}{e F}+s \cdot \coth eFs {]}_{\lambda \nu} \ ,
\\
&& O_{\lambda \nu}=2 R(u) I^{-1}(s) \ ,
\\
&&\int_0^s du  \ O_{\nu \lambda}=g_{\nu \lambda} \ .
 \eea

The result of eq.[\ref{qm}] is
 \bea
\label{undertrace}
\nonumber
&&(x^{\prime}(s) | U_I(s) | x^{\prime\prime} (0))=  \\
&& \ \ (x^{\prime}(s) | x^{\prime\prime}(0)) (-ig_a\gamma_5
A_\mu^5)\left[-s k_\mu - i\tilde\sigma_{\mu\lambda}{\cal
D}_{\lambda\nu}k_\nu + i\tilde\sigma_{\mu
\nu}(x^{\prime}-x^{\prime \prime})_\nu\right] \ ,
 \eea

Using the property of the trace we can replace
$\tilde\sigma_{\mu \nu} \to \sigma_{\mu \nu}$.

We make simplifications
\bea
&&{\cal D}_{\lambda\nu} e^{-l(s)} = \frac{2}{e}\frac{ d e^{-l(s)}}
{dF_{\lambda\nu}} \ ,
%= \frac{2}{e} \frac{ d C_1({\cal F},{\cal G})}{dF_{\lambda\nu}}
\\
&&\gamma_5\sigma_{\mu\lambda}e^{\frac{i}{2}e\sigma F s}=
\frac{2}{es}\frac{ d (-i\gamma_5e^{\frac{i}{2}e\sigma F
s})}{dF_{\mu \lambda}} \ .
%=\frac{2}{es}\frac{ d
%C_2({\cal F},{\cal G})}{dF_{\lambda\nu}}
 \eea

We define two functions
 \bea
C_1({\cal F},{\cal G}) &\equiv& e^{-l(s)}  \ ,
\\
C_2({\cal F},{\cal G}) &\equiv& -i Tr \gamma_5 e^{\frac{i}{2}e\sigma F
s} \ ,
\eea
where
\bea
&&{\cal F} = \frac{1}{4}F_{\mu \nu} F_{\mu \nu} \ ,
\\
&&{\cal G} = \frac{1}{4}F_{\mu \nu} \tilde F_{\mu \nu} \ ,
\\
&&F_{\mu \nu} =\partial_\mu A_\nu - \partial_\nu A_\mu \ ,
\\
&& \tilde F_{\mu \nu} = \frac{1}{2}e^{\mu \nu \rho \sigma}F_{\rho \sigma}
 \ ,  \ \ \ \ \ \ e^{0123}=1 \ .
 \eea

Now the trace of the eq.[\ref{qm}] will have a form
 \bea
\label{trqm}
&&\hspace{-1cm} Tr(x^{\prime}(s)
|U_I(s)|x^{\prime\prime} (0))= \frac{-i}{s^2(4\pi)^2}
\Phi(x^{\prime},x^{\prime \prime}) e^{(x^{\prime}-x^{\prime
\prime})E(s)(x^{\prime}-x^{\prime \prime})}
\nonumber \\
&&(g_aA_\mu^5) \{ -s C_1 C_2 k_\mu
-\frac{4}{e^2s}\frac{dC_1}{d F_{\mu \lambda}}\frac{dC_2}{d F_{\lambda
\nu}}k_\nu
+ \frac{2}{e s}C_1\frac{dC_2}{dF_{\mu \nu}}(x^{\prime}-x^{\prime
\prime})_\nu \} \ .
 \eea

Making use of
 \bea
&&\frac{d}{dF_{\lambda\nu}} = \frac{1}{2}\frac{d}{d {\cal
F}}F_{\lambda\nu}+ \frac{1}{2}\frac{d}{d {\cal G}} \tilde
F_{\lambda\nu} \ ,
\\
&&F_{\mu \lambda} \tilde F_{\lambda \nu }=-g_{\mu \nu}{\cal G} \ ,
\\
&&\tilde F_{\mu \lambda} \tilde F_{\lambda \nu } -
F_{\mu \lambda} F_{\lambda \nu }=2g_{\mu \nu}{\cal F} \ ,
 \eea
we can write
 \bea
\nonumber 4\frac{dC_1}{d F_{\mu \lambda}}\frac{dC_2}{d
F_{\lambda\nu}}&=& \left[ \frac{dC_1}{d {\cal F}}\frac{dC_2}{d
{\cal F}}+ \frac{dC_1}{d {\cal G}}\frac{dC_2}{d {\cal G}}\right]
F_{\mu \lambda}F_{\lambda \nu}
\\
&+&g_{\mu \nu}\left[-{\cal G} \left(\frac{dC_1}{d{\cal
F}}\frac{dC_2}{d{\cal G}} +\frac{dC_1}{d{\cal
G}}\frac{dC_2}{d{\cal F}}\right)+2{\cal F}\frac{dC_1}{d{\cal
G}}\frac{dC_2}{d{\cal G}} \right] \eea

For the first two terms in the braces of the
eq.[\ref{trqm}]
we may put $x^{\prime}-x^{\prime \prime}=0$.

Now we'll show that the singular part of the last term in the breces of the 
eq.[\ref{trqm}] gives triangle anomaly and non singular parts give
additional contributions to the effective Lagrangian.

We integrate the last term in eq.[\ref{trqm}]  by parts using the
property
 \beq \label{gaug}
 eF_{\lambda\nu}(x^{\prime}-x^{\prime
\prime})_\nu \Phi(x^{\prime},x^{\prime \prime})=
-i\partial_\lambda \Phi(x^{\prime},x^{\prime \prime}) \ .
 \eeq

By using the identities
 \bea
&&\frac{d C_2}{dF_{\lambda\nu}} =
 \frac{1}{2}\frac{d C_2}{d {\cal F}}F_{\lambda\nu}+
\frac{1}{2}\frac{d C_2}{d {\cal G}} \tilde
 F_{\lambda\nu} \ ,
 \\ &&
\tilde F_{\mu \nu} = \frac{\tilde F_{\mu \nu}{\cal G}}{{\cal G}}=
\frac{-(\tilde F\tilde F F)_{\mu \nu}}{{\cal G}}= \frac{- (F_{\mu
\lambda} F_{\lambda \delta} +2g_{\mu \delta} {\cal F})}{{\cal
G}}F_{\delta \nu}
 \eea
and defining
 \beq
H = \Phi(x^{\prime},x^{\prime \prime})
e^{(x^{\prime}-x^{\prime \prime})E(s)(x^{\prime}-x^{\prime \prime})}
C_1\frac{dC_2}{d {\cal G}}
\eeq
we can develop
\bea
&&H\tilde F_{\mu \nu}
(x^{\prime}-x^{\prime \prime})_\nu
\\
\nonumber
&&= \left(H-H\mid_{{\cal G}=0}\right)
\frac{- (F_{\mu \lambda} F_{\lambda \delta} +2g_{\mu\delta}
{\cal F})}{{\cal G}}F_{\delta \nu}(x^{\prime}-x^{\prime\prime})_\nu
+H\mid_{{\cal G}=0}\tilde F_{\mu \nu}
(x^{\prime}-x^{\prime \prime})_\nu
\\
\nonumber
&&\to \left(H-H\mid_{{\cal G}=0}\right)
\frac{F_{\mu \lambda} F_{\lambda \delta}
+2g_{\mu\delta}
{\cal F}}{e {\cal G}}k_\delta
+H\mid_{{\cal G}=0}\tilde F_{\mu \nu}
(x^{\prime}-x^{\prime \prime})_\nu \ .
 \eea

In other words, before integrating by parts we extract the
singular part
 \bea
 C_1 \frac{d C_2}{d {\cal G}}\mid_{{\cal G}=0}=4e^2s^2 \ .
\\
\eea

Finally collecting all terms and changing integration variables
$s \to -is$ we get effective Lagrangian 

 \beq \label{eflan}
{\cal L } =  \frac{i g_a m^2}{32 \pi^2 }A_\mu^5 k_\nu \left\{ Ag_{\mu
\nu}+\frac{e^2}{m^4} B F_{\mu \lambda} F_{\lambda \nu} \right\}
+  anomaly \hspace{3mm}term
 \ ,
 \eeq
where
 \bea
\nonumber
A&=&\frac{1}{m^2}\int_0^\infty
\frac{ds}{s^2}e^{-m^2s}\Big\{-C_1C_2+\frac{1}{e^2s^2}
\Big[-{\cal G}
\left(\frac{dC_1}{d{\cal F}}\frac{dC_2}{d{\cal G}}
+\frac{dC_1}{d{\cal G}}\frac{dC_2}{d{\cal F}}\right)
\\
&&+2{\cal F}\frac{dC_1}{d{\cal G}}\frac{dC_2}{d{\cal G}}
+C_1\frac{dC_2}{d{\cal F}}
-\frac{2{\cal F}}{{\cal G}}
\left(C_1\frac{dC_2}{d{\cal G}}+4e^2s^2 \right)\Big]\Big\}
 \ ,
\\
B&=&\frac{m^2}{e^4}\int_0^\infty
\frac{ds}{s^4}e^{-m^2s}\Big\{
\frac{d C_1}{d {\cal F}}
\frac{d C_2}{d {\cal F}}
%\nonumber
%\\
%&&
+\left(
\frac{d C_1}{d {\cal G}}-
\frac{  C_1}{  {\cal G}} \right)
\frac{d C_2}{d {\cal G}}-\frac{4 s^2}{{\cal G}}
\Big\}
 \ .
 \eea
Here
\bea
C_1({\cal F},{\cal G}) &=&
\frac{2ie^2s^2 {\cal G}}
{\cosh (es\sqrt{2({\cal F}+i{\cal G})})-
\cosh (es\sqrt{2({\cal F}-i{\cal G})})}
 \ ,
\\
C_2({\cal F},{\cal G}) &=& 2 i \left(
\cosh (es\sqrt{2({\cal F}+i{\cal G})})-
\cosh (es\sqrt{2({\cal F}-i{\cal G})}) \right)
 \ .
 \eea

The first terms of the Taylor expansion of $B$ have a form
 \beq
 B=\frac{8}{15}\frac{e^2}{m^4}{\cal G}-\frac{128}{63}\frac{e^4}{m^8}
{\cal G}{\cal F} +\frac{256}{15}\frac{e^6}{m^{12}}{\cal F}^2{\cal
G} +\frac{64}{15} \frac{e^6}{m^{12}}{\cal G}^3 +\cdots
 \eeq

The effective Lagrangian [\ref{eflan}]  has very simple Lorenz
invariant form. Along with the anomaly term, it also contains two  Lorenz
invariant terms.

 As it is well known from  Euler Heisenberg Lagrangian (for vector currents (not-axial)) one can get  amplitudes for n real photons interactions at low energies (for instance photon photon scattering, photon splitting in the magnetic field etc). Those amplitudes are calculated via replacing $ F_{\mu \nu} \to F_{\mu \nu} + f^1_{\mu
\nu}+f^2_{\mu \nu}+ \cdots$ ( where $f^i_{\mu \nu}$ are real
photons and $F_{\mu \nu}$ is the constant external electromagnetic
field) in the effective  Euler Heisenberg Lagrangian. It is interesting to note that the effective Lagrangian (67) is not possible
to use for calculating amplitudes of low energy processes which include real
photons.

The photon- Z boson mixing  in the external magnetic field
was calculated in \cite{Ioannisian:1996pn}. The amplitude
has more complicated Lorenz invariant terms, than one can expect from our effective Lagrangian (67). Those additional terms include the 
derivative of the real photon field ( $\partial_\lambda f_{\mu
\nu} \ , \ \partial_\mu f_{\mu \nu} $ ). It is impossible to
relate that derivative to the Z boson field via integration
by parts of the photon- Z boson mixing amplitude. Therefore it is impossible to restore our effective Lagrangian (with only two Lorenz invariant terms) from the photon- Z boson mixing amplitude. 

The effective Euler Heisenberg Lagrangian with one axial-vector field must include terms with derivatives of external electromagnetic field. And amplitude is not vanishing even when energy-momentum is not transferring from axial-vector field to the external electromagnetic field (thanks to it's derivatives). For instance there will be non vanishing forward scattering amplitude of neutrinos in the external varying electromagnetic field (as a result index of refraction of neutrinos will departure from the unity).

\vspace {0.3cm}

Now we turn to the anomaly term. We note that

\beq
E(s)_{\mu \nu}= E_1g_{\mu \nu}+e^2 E_2 F_{\mu
\lambda} F_{\lambda \nu}
 \eeq

In the limit ${\cal G} \to 0$ the functions $E_1(s)$ and $E_2(s)$ have
forms:

for ${\cal F}\ge 0$
\bea
&&E_1= \frac{i}{4s}
\\
&&E_2=-\frac{i(-1+es\sqrt{2 {\cal F}}\cot{(es\sqrt{2 {\cal F}})})}
{8 s e^2{\cal F}} \simeq O(s)
\eea

for ${\cal F}\le 0$
\bea
&&E_1= \frac{ies\sqrt{2 {\cal F}}\cot{(es\sqrt{2 {\cal F}})}}{4s} \simeq
 \frac{i}{4s}+O(s)
\\
&&E_2=\frac{i(-1+es\sqrt{2 {\cal F}}\cot{(es\sqrt{2 {\cal F}})})}
{8 s e^2{\cal F}} \simeq O(s)
 \eea

When $x^{\prime} \simeq x^{\prime \prime}$ the integral can be
taken
 \beq
\int_0^\infty \frac{ds}{s^2}e^{-im^2s}
e^{(x^{\prime}-x^{\prime \prime}) E(s)(x^{\prime}-x^{\prime \prime})}=
\int_0^\infty \frac{ds}{s^2}e^{\frac{i(x^{\prime}-x^{\prime
\prime})^2}{4s}}=
\frac{-i4}{(x^{\prime}-x^{\prime \prime})^2}
 \eeq

Therefore the anomaly term is equal to
 \beq
-g_aA_\mu^5\frac{ie}{2\pi^2}\Phi(x^{\prime},x^{\prime \prime})
\tilde F_{\mu \nu}(x^{\prime}-x^{\prime \prime})_\nu(x^{\prime}-x^{\prime
\prime})^{-2}
\eeq

Now by replacing $g_aA_\mu^5 \to \partial_\mu $  and using eq. [\ref{gaug}] we get

 \beq
 \frac{2\alpha}{\pi}{\cal G} \ .
 \eeq

At the end we want to mention, that we may get the same result for
effective Lagrangian  by calculating axial vector current in the
presence of the external constant electromagnetic field
 \bea
<j_\mu^5> &=& i g_a Tr \gamma_\mu \gamma_5 G(x^\prime,x^{\prime
\prime})
\\
\nonumber
&=&\frac{g_a}{2}\int_0^\infty ds e^{-im^2s} Tr \gamma_5 {[}g_{\mu\nu}
\left(x(s)^\prime |\Pi_\nu(s)-\Pi_\nu(0) |x(0)^{\prime \prime}\right)
\\
&+&i\sigma_{\mu \nu}\left(x(s)^\prime |\Pi_\nu(s)+\Pi_\nu(0) |x(0)^{\prime
\prime}\right){]}
 \ .
 \eea

Here
 \bea
\left(x(s)^\prime |\Pi(s) | x(0)^{\prime\prime}\right)=
\frac{1}{2}\left[eF\coth(eFs)+eF\right](x^\prime -x^{\prime\prime})
\left(x(s)^\prime | x(0)^{\prime\prime}\right)
 \ ,
 \\
\left(x(s)^\prime |\Pi(0) | x(0)^{\prime\prime}\right)=
\frac{1}{2}\left[eF\coth(eFs)-eF\right](x^\prime -x^{\prime\prime})
\left(x(s)^\prime | x(0)^{\prime\prime}\right)
 \ .
 \eea

Finally
 \bea
\nonumber <j_\mu^5> &=& \frac{g_a}{2}\int_0^\infty ds e^{-im^2s}
Tr \gamma_5 {[} eF_{\mu \nu}(x^\prime -x^{\prime\prime})_\nu
\\
&&+i\sigma_{\mu \nu}\left[eF\coth(eFs)\right]_{\nu \lambda}(x^\prime
-x^{\prime\prime})_\lambda {]}
\left(x(s)^\prime |x(0)^{\prime \prime}\right)
\eea

After integrating this result by parts, using eq. [\ref{gaug}], we
get under the trace the same expression as in
eq.[\ref{undertrace}].

\vspace{1cm}

We are calculated the expectation value of the axial-vector current induced by the vacuum polarization effect of the Dirac field in constant external electromagnetic field.
In calculations we use Schwinger's proper time method. 
The effective Lagrangian, eq.[\ref{eflan}], has very simple Lorenz
invariant form. Along with the anomaly term, it also contains two  Lorenz
invariant terms. 
The result is compared with our previous
calculation of the photon - Z boson mixing in the magnetic field.

\end{document}